\documentclass[prb,superscriptaddress,preprint,endfloats,nopacs]{revtex4}

\usepackage{amsmath, amsthm, amssymb, pifont, wasysym}
\usepackage{graphicx}
\usepackage{bm} 
\usepackage{pifont}
\usepackage{dcolumn} 
\begin{document}
	\title{
		Quantum transport observed in films of magnetic topological semimetal EuSb$_2$
	}
	
	\author{Mizuki Ohno}
	\affiliation{Department of Applied Physics, University of Tokyo, Tokyo 113-8656, Japan}
	\affiliation{Quantum-Phase Electronics Center (QPEC), University of Tokyo, Tokyo 113-8656, Japan}
	\author{Masaki Uchida}
	\email[Author to whom correspondence should be addressed: ]{m.uchida@phys.titech.ac.jp}
	\affiliation{Department of Applied Physics, University of Tokyo, Tokyo 113-8656, Japan}
	\affiliation{Quantum-Phase Electronics Center (QPEC), University of Tokyo, Tokyo 113-8656, Japan}
	\affiliation{PRESTO, Japan Science and Technology Agency (JST), Tokyo 102-0076, Japan}
	\affiliation{Department of Physics, Tokyo Institute of Technology, Tokyo 152-8550, Japan}
	\author{Ryosuke Kurihara}
	\affiliation{The Institute for Solid State Physics (ISSP), The University of Tokyo, Kashiwa, Chiba 277-8581, Japan}
	\affiliation{RIKEN Center for Emergent Matter Science (CEMS), Wako 351-0198, Japan}
	\author{Susumu Minami}
	\affiliation{Department of Physics, University of Tokyo, Tokyo 113-8656, Japan}
	\author{Yusuke Nakazawa}
	\affiliation{Department of Applied Physics, University of Tokyo, Tokyo 113-8656, Japan}
	\affiliation{Quantum-Phase Electronics Center (QPEC), University of Tokyo, Tokyo 113-8656, Japan}
	\author{Shin Sato}
	\affiliation{Department of Applied Physics, University of Tokyo, Tokyo 113-8656, Japan}
	\affiliation{Quantum-Phase Electronics Center (QPEC), University of Tokyo, Tokyo 113-8656, Japan}
	\author{Markus Kriener}
	\affiliation{RIKEN Center for Emergent Matter Science (CEMS), Wako 351-0198, Japan}
	\author{Motoaki Hirayama}
	\affiliation{Department of Applied Physics, University of Tokyo, Tokyo 113-8656, Japan}
	\affiliation{RIKEN Center for Emergent Matter Science (CEMS), Wako 351-0198, Japan}
	\author{Atsushi Miyake}
	\affiliation{The Institute for Solid State Physics (ISSP), The University of Tokyo, Kashiwa, Chiba 277-8581, Japan}
	\author{Yasujiro Taguchi}
	\affiliation{RIKEN Center for Emergent Matter Science (CEMS), Wako 351-0198, Japan}
	\author{Ryotaro Arita}
	\affiliation{Department of Applied Physics, University of Tokyo, Tokyo 113-8656, Japan}
	\affiliation{RIKEN Center for Emergent Matter Science (CEMS), Wako 351-0198, Japan}
	\author{Masashi Tokunaga}
	\affiliation{The Institute for Solid State Physics (ISSP), The University of Tokyo, Kashiwa, Chiba 277-8581, Japan}
	\affiliation{RIKEN Center for Emergent Matter Science (CEMS), Wako 351-0198, Japan}
	\author{Masashi Kawasaki}
	\affiliation{Department of Applied Physics, University of Tokyo, Tokyo 113-8656, Japan}
	\affiliation{Quantum-Phase Electronics Center (QPEC), University of Tokyo, Tokyo 113-8656, Japan}
	\affiliation{RIKEN Center for Emergent Matter Science (CEMS), Wako 351-0198, Japan}
	
	
	\begin{abstract}
		We report fabrication of EuSb$_2$ single-crystalline films and investigation of their quantum transport.
		First-principles calculations demonstrate that EuSb$_2$ is a magnetic topological nodal-line semimetal protected by nonsymmorphic symmetry.
		Observed Shubnikov-de Haas oscillations with multiple frequency components exhibit small effective masses and two-dimensional field-angle dependence even in a 250~nm thick film, further suggesting possible contributions of surface states.
		This finding of the high-mobility magnetic topological semimetal will trigger further investigation of exotic quantum transport phenomena by controlling magnetic order in topological semimetal films.
	\end{abstract}
	
	\maketitle
	\section{INTRODUCTION}
	Topological semimetals are characterized by nontrivial band crossings at nodal points or nodal lines in the momentum space \cite{Fang2016,Burkov2011,Lu2017b,Wang2017f,Armitage2018}.
	In topological nodal-point semimetals, Dirac or Weyl semimetals, these gapless points give rise to exotic quantum transport phenomena \cite{Lu2017b,Wang2017f}.
	Under magnetic fields, for example, Fermi-arcs of the surface state merge with the bulk state and form so-called Weyl orbits \cite{Potter2014a, Wang2017i}, as recently observed in its quantum Hall state \cite{Zhang2017a,Lin2019,Zhang2019e,Nishihaya2019}.
	
	Topological nodal line semi-metals (TNLSMs) are characterized by the Zak phase \cite{Zak1989} in loops encircling the nodal lines.
	It has been theoretically suggested that this topological feature appears in bulk \cite{Oroszlany2018a,Yang2018} 
	and surface \cite{Chan2016, Bian2016c, Yamakage2016, Hirayama2017, Hirayama2018} quantum oscillations.
	In this context, it is important to prepare high-mobility films of TNLSMs.
	On the other hand, many compounds which possess nodal lines without spin-orbit coupling (SOC) turn into topological insulators or nodal point semimetals when considering SOC \cite{Fang2015, Kobayashi2017}.
	In order to protect these nodal lines from gapping by strong SOC, nonsymmorphic symmetry is required \cite{Fang2015, Kobayashi2017, Zhao2016a}.
	
	In general, time-reversal symmetry breaking destroys the robustness of nodal lines \cite{Okugawa2017}, while they remain stable only when the magnetic ordering still respects space group symmetries \cite{Bzdusek2016, Wang2017a}.
	Therefore, there are very few reports about intrinsic magnetic TNLSMs under SOC (single-layer GdAg$_2$ \cite{Ormaza2016, Feng2019} and GdSbTe \cite{Hosen2018, Sankar2019}).
	Many of the nodes are slightly gapped on the entire lines (Fe$_3$GeTe$_2$ \cite{Chen2013, Kim2018} and Fe$_3X$ ($X=$Ga, Al) \cite{Sakai2020b}) or except the Weyl points (Co$_3$Sn$_2$S$_2$ \cite{Liu2018a} and Co$_2$MnGa \cite{Chang2017,Sakai2018, Belopolski2019a}) under SOC.
	Moreover, quantum oscillations have not been observed in all of these compounds.
	Hence, to identify new magnetic TNLSMs is strongly called for exploring unique quantum transport phenomena.
	
	CaSb$_2$ is a novel TNLSM where nodal lines are protected by two-fold screw symmetry along $b$ axis even in the presence of SOC \cite{Funada2019}.
	Recently, superconductivity has been also observed in polycrystalline samples \cite{Ikeda2020a}.
	In EuSb$_2$, on the other hand, large magnetic moments are expected to be introduced while keeping the unique crystal structure.
	Although antiferromagnetic (AFM) ordering of nearly Eu$^{2+}$ spins has been investigated in the 1980s \cite{Hulliger1978, Niggli1984}, electronic structures and transport properties have long been unknown for EuSb$_2$.
	Here we demonstrate that EuSb$_2$ is a new magnetic TNLSM protected by nonsymmorphic symmetry of the space group $P2_1$/$m$.
	We fabricate single-crystalline EuSb$_2$ films by molecular beam epitaxy and investigate their magnetotransport.
	Observed Shubnikov-de Haas (SdH) oscillations with multiple frequency components indicate small effective masses and two-dimensional (2D) field-angle dependence characteristics in topological semimetals.
	
	\section{EXPERIMENTAL METHODS}
	EuSb$_2$ epitaxial films were grown on single-crystalline (11$\bar{2}$0) Al${_{2}}$O${_{3}}$ substrates in an Epiquest RC1100 chamber. 
	The molecular beams were simultaneously provided from conventional Knudsen cells containing 3N Eu and 6N Sb.
	The growth temperature was set at 850~${^\circ}$C, and the beam equivalent pressures, measured by an ionization gauge, were set to 1${\times}$10${^{-5}}$~Pa for Eu and 4.0${\times}$10${^{-4}}$~Pa for Sb, considering the Eu-Sb binary phase diagram \cite{Abdusalyamova2011}.
	The film thicknesses were typically set at 250~nm for structural characterization and magnetotransport measurements, and 750~nm for magnetization measurements.
	The growth rate was about 0.3~\AA/s.
	High-field resistivity and magnetization measurements were performed at the International MegaGauss Science Laboratory in the Institute for Solid State Physics at the University of Tokyo.
	Resistivity up to 58~T was measured by a standard four-probe method for 200~$\mu$m-width multi-terminal Hall bars, using a non-destructive pulsed magnet \cite{Uchida2017}.
	Temperature dependence was also measured using a Quantum Design Physical Properties Measurement System.
	Magnetization up to 27~T was measured by a conventional induction method with coaxial pickup coil \cite{Yoshida2009a}.
	Temperature dependence was also measured using a superconducting quantum interference device magnetometer in a Quantum Design Magnetic Property Measurement System.
	First-principles calculations of the band structure with SOC were performed by using the VASP package \cite{Kresse1996, Kresse1996a, Kresse1999}, with the experimentally determined lattice constants \cite{Hulliger1978}.
	The generalized gradient approximation of Perdew-Burke-Ernzerhof was adopted for the exchange-correlation functional \cite{Perdew1996}.
	$16\times16\times10$~$k$-point mesh with Monkhorst-Pack scheme \cite{Monkhorst1976} was used for the Brillouin zone sampling of the primitive cell and Gaussian smearing with a width of $0.02$~eV was applied.
	
	\section{RESULTS AND DISCUSSION}
	\begin{figure}
		\begin{center}
			\includegraphics*[bb=0 0 576 459,width=13.5cm]{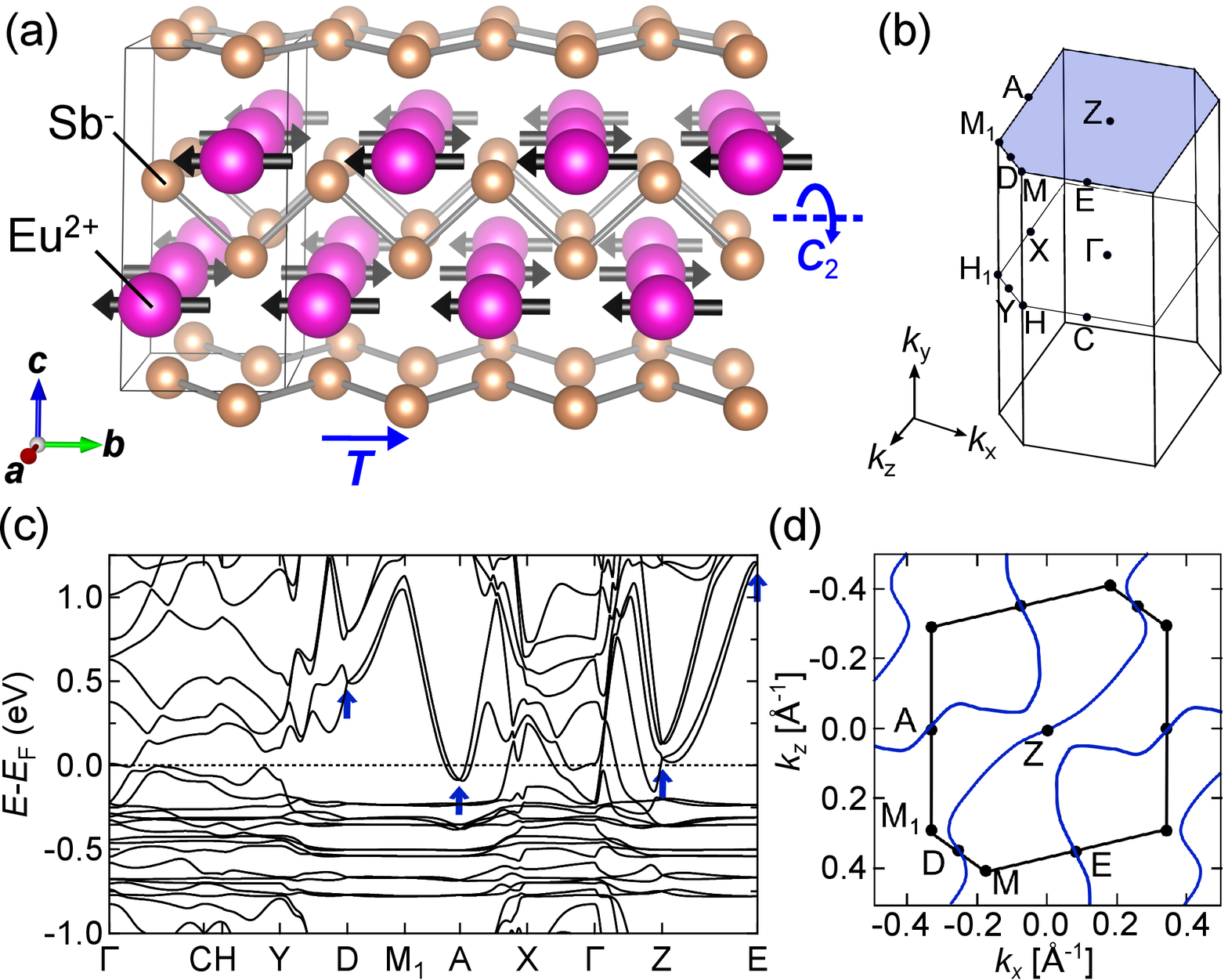}
			\caption{
				(a) Crystal structure of EuSb$_2$ with AFM ordering of Eu$^{2+}$ spins at zero field.
				The solid lines indicate the unit cell, which is doubled along the $a$ axis by the AFM ordering.
				The screw symmetry that protects the nodal lines is represented in combination with rotational ($C_2$) and translational ($T$) operations.
				(b) Brillouin zone of EuSb$_2$. The $\Gamma$-Z line is parallel to the $k_y$ axis.
				(c) Band structure of EuSb$_2$ calculated for the AFM phase including SOC.
				Band crossings at the high-symmetry points, which are protected by the two-fold screw symmetry, are denoted by blue arrows.
				(d) Line nodes appearing on the $k_y=\pi /b$ plane in the presence of SOC.
			}
			\label{fig1}
		\end{center}
	\end{figure}
	Figure~1(a) shows the crystal structure of EuSb$_2$.
	This structure is characterized by Sb zigzag chains running parallel to the $b$ axis and mirror planes in the $ac$ plane.
	This Sb chain forms Sb$_n$$^{n-}$ and the valence of Eu becomes divalent (Eu$^{2+}$).
	These Eu spin moments are oriented along the [010] direction and antiferromagnetically ordered with a propagation vector of $(1/2,0,0)$, corresponding to the magnetic space group $P_a2_1/m$ \cite{Niggli1984}.
	Therefore, this AFM ordering does not break the two-fold screw symmetry ($2_1/m$) which protects the nodal lines on the $k_y=\pi/b$ plane.
	It is thus expected that the nodal lines stably exist in the AFM phase.
	
	The Brillouin zone and band structure calculated for the AFM phase are presented in Figs.~1(b) and 1(c).
	While flat bands located between -0.3 and -0.7~eV are attributed to the Eu 4$f$ orbitals, dispersive bands crossing $E_\mathrm{F}$ originate from the Sb 5$p$ ones, resembling TNLSM CaSb$_2$.
	There are band crossing points at the high-symmetry points Z, A, E, and D as in CaSb$_2$ \cite{Funada2019}, and which are connected by nodal lines on the $k_y=\pi /b$ plane, as shown in Fig.~1(d).
	This indicates that EuSb$_2$ is a magnetic TNLSM even in the presence of SOC.
	
	\begin{figure}
		\begin{center}
			\includegraphics*[bb=0 0 729 451,width=15cm]{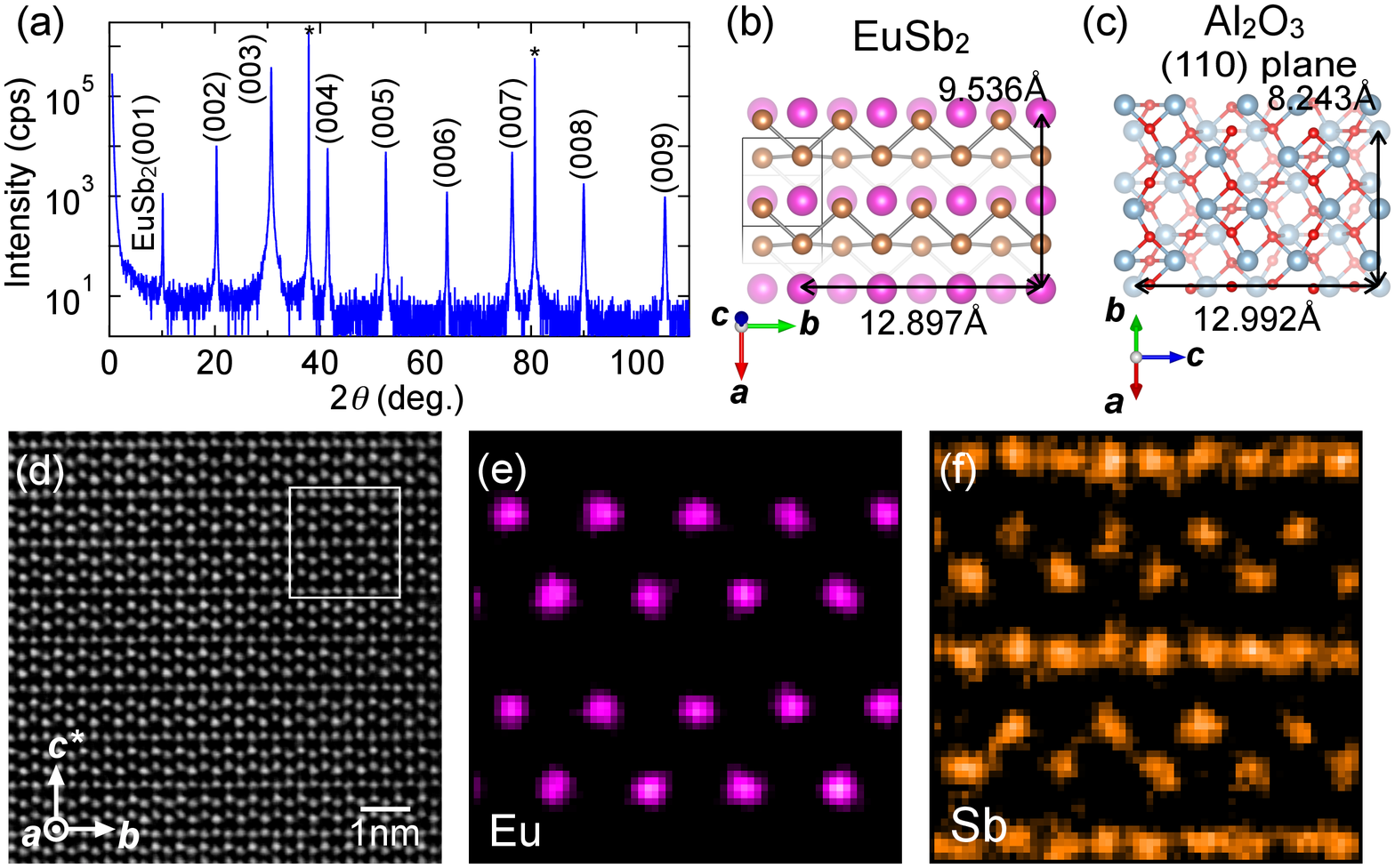}
			\caption{
				(a) XRD ${\theta}$-${2\theta}$ scan of a EuSb$_2$ film grown on Al${_{2}}$O${_{3}}$ (11$\bar{2}$0) substrate. Al$_2$O$_3$ substrate peaks are marked with an asterisk.
				Epitaxial relation between (b) the EuSb$_2$ film and (c) the Al${_{2}}$O${_{3}}$ substrate viewed along the out-of-plane direction.
				(d) Cross-sectional image of the EuSb$_2$ film, taken by high-angle annular dark-field scanning transmission electron microscopy. Energy dispersive X-ray spectrometry map for Eu $L$ (e) and Sb $L$ (f) edges in the boxed region in (d).
				The Sb zigzag chains along the $b$ axis are atomically resolved.
			}
			\label{fig2}
		\end{center}
	\end{figure} 
	X-ray diffraction (XRD) ${\theta}$-${2\theta}$ scan in Fig.~2(a) reveals sharp reflections from the (001) EuSb$_2$ lattice plane without any impurity phases.
	Figures.~2(b) and 2(c) show the epitaxial relation viewed along the out-of-plane $c^\ast$ axis, with a mismatch of 0.7~\% between the $c$ axis of the (11$\bar{2}$0) Al${_{2}}$O${_3}$ plane and the $b$ axis of the (001) EuSb$_2$ plane.
	The $c$ axis of EuSb$_2$ is tilted by approximately $13$~deg. from the $c^\ast$ axis perpendicular to the $a$-$b$ plane.
	Figure~2(d) shows a cross-section image of the EuSb$_2$ film, taken by high-angle annular dark-field scanning transmission electron microscopy.
	The periodic arrangement of EuSb$_2$ is clearly confirmed which corresponds to the EuSb$_2$ crystal structure determined by single-crystal XRD \cite{Hulliger1978}.
	Elemental maps are taken by energy dispersive X-ray spectrometry as shown in Figs.~2(e) and 2(f).
	The Sb zigzag chains along the $b$ axis are atomically resolved.
	Further XRD characterization was also performed to confirm high crystallinity and in-plane orientation (see Supplementary Materials \cite{Supplement}).
	
	\begin{figure}
		\begin{center}
			\includegraphics*[bb=1 2 678 571,width=15cm]{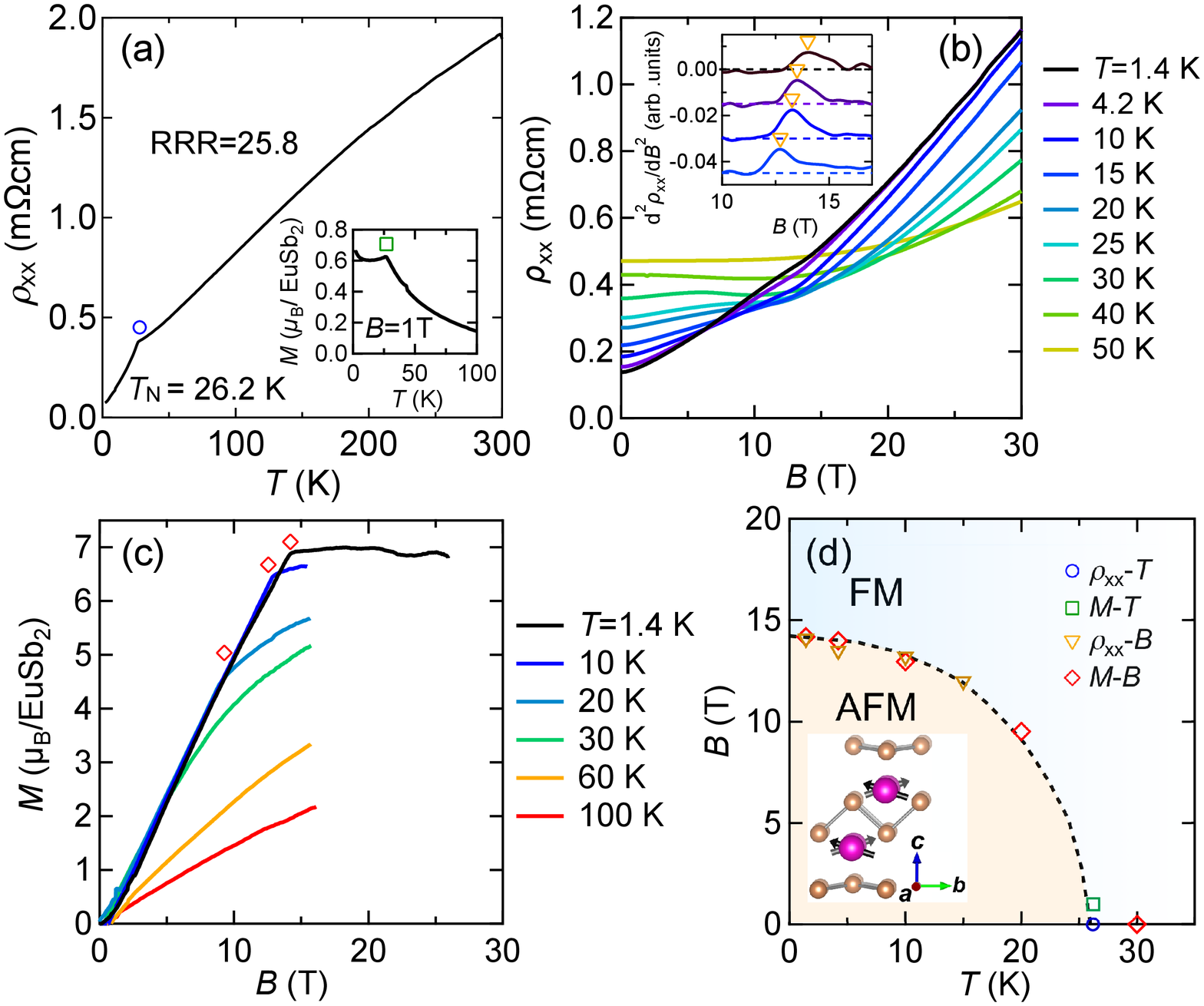}
			\caption{
				(a) Temperature dependence of the resistivity ${\rho_{\mathrm{xx}}}$ measured at zero field.
				The inset shows temperature dependence of the magnetization $M$, taken with applying an out-of-plane magnetic field of $1.0$~T.
				Both ${\rho_{\mathrm{xx}}}$ and $M$ exhibit a clear kink at the N\'{e}el temperature ${T_{\mathrm{N}}}$ of $26.2$~K, as marked by a circle and a square respectively.
				(b) ${\rho_{\mathrm{xx}}}$ taken with sweeping the out-of-plane magnetic field at various temperatures.
				The inset shows second derivative of $\rho_\mathrm{xx}$, where the saturation field is more clearly confirmed as marked by a triangle.
				(c) Magnetization curve taken at various temperatures, where the saturation field is marked by a diamond.
				(d) Magnetic phase diagram obtained for out-of-plane magnetic fields ($B\parallel c^\ast$).
				The dashed  curve represents the boundary between the AFM and FM phases.
				As shown in the inset, the antiferromagnetically ordered Eu$^{2+}$ spins along the $b$ axis at zero field are gradually canted toward the $c^\ast$ axis upon increasing the magnetic field.}
			\label{fig3}
		\end{center}
	\end{figure}
	Figure~3(a) summarizes temperature dependence of the resistivity.
	The resistivity exhibits metallic behavior down to $2$~K with a residual resistivity ratio (RRR$=\rho_\mathrm{xx}(300\,\mathrm{K})/\rho_\mathrm{xx}(2\,\mathrm{K})$) of $26$, reflecting high quality of the film.
	A clear kink observed at 26.2~K corresponds to the N\'{e}el temperature ($T_{\mathrm{N}}$) \cite{Hulliger1978}, which is also seen in temperature dependence of the magnetization (inset).
	The resistivity decreases rapidly with decreasing temperature below $T_\mathrm{N}$ probably due to suppression of the magnetic fluctuations.
	Figure~3(b) presents out-of-plane magnetic field sweeps of the resistivity up to 58~T at various temperatures.
	The resistivity at the base temperature of 1.4~K exhibits a cusp-like behavior at 14.1~T, which is more clearly seen in its second derivative (inset).
	This saturation field shifts to the lower values with increasing temperature and then disappears above ${T_{\mathrm{N}}}$.
	Figure~3(c) shows magnetization curves measured at various temperatures.
	The magnetization taken at $1.4$~K increases approximately linearly with field and then saturates at about 7$\mu_B$, indicating transition from the AFM phase at low fields to the forced ferromagnetic (FM) one.
	This behavior is consistent with previous magnetization measurements \cite{Hulliger1978}.
	Below ${T_{\mathrm{N}}}$, the saturation field shifts to lower values with increasing temperature, as confirmed in the resistivity change.
	Above ${T_{\mathrm{N}}}$, the kink indicating saturation disappears and the magnetization curve follows the Brillouin function.
	Therefore, the kink indicates the phase transition from AFM to FM phase, where Eu$^{2+}$ $S=7/2$ spins are polarized along the ${c^\ast}$ direction by the magnetic field.
	
	Figure~3(d) summarizes a $B$-$T$ phase diagram as determined by the above transport and magnetization measurements.
	Upon increasing the out-of-plane magnetic field, the Eu$^{2+}$ spin moments, which are initially oriented along the $b$ axis at zero field, are gradually canted toward the $c^\ast$ axis without spin-flop transition.
	For the in-plane field, on the other band, the magnetization below ${T_{\mathrm{N}}}$ exhibits a spin-flop transition at $2.4$~T, corresponding to reorientation of the spin moments perpendicular to the applied field (see Supplementary Materials \cite{Supplement}).
	EuSb$_2$ exhibits large positive magnetoresistance both in the AFM and FM phases, similar to nonmagnetic TNLSMs (such as ZiSiS \cite{Sankar2017} and CaSb$_2$ \cite{Funada2019}).
	In the FM phase, the Eu$^{2+}$ localized spins are polarized along the magnetic-field direction and electron scattering by the localized spins is expected to be further suppressed.
	
	\begin{figure}
		\begin{center}
			\includegraphics*[bb= 1 1 695 576,width=15cm]{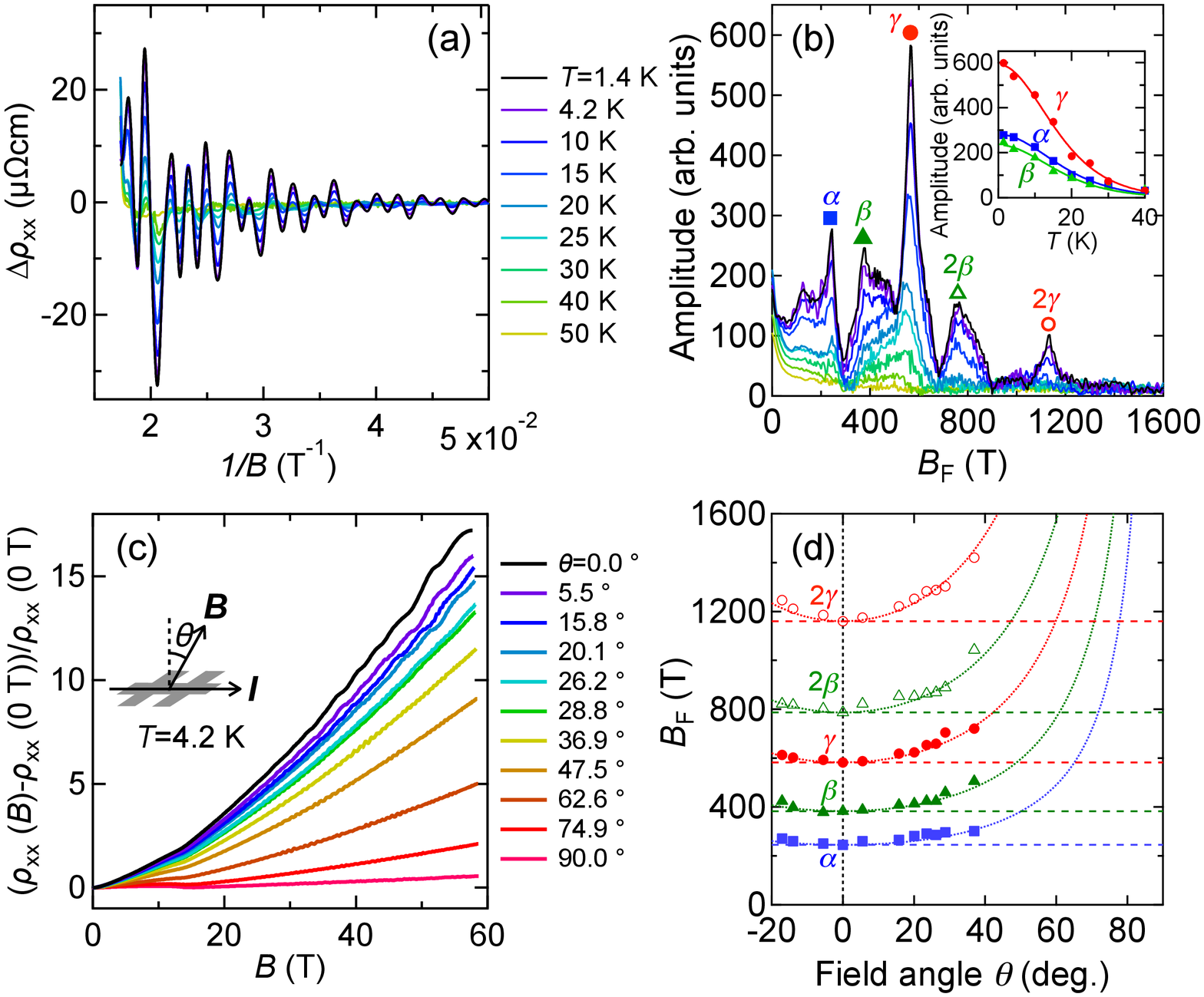}
			\caption{
				(a) Temperature dependence of the SdH oscillation $\Delta \rho_{\mathrm{xx}}$, plotted against $1/B$ after subtracting a smooth background from $\rho_\mathrm{xx}$.
				(b) FFT spectra of the SdH oscillation $\Delta \rho_{\mathrm{xx}}$ for the same set of temperatures.
				As shown in the inset, temperature dependence of the FFT amplitudes of $\alpha, \beta$, and $\gamma$ peaks is well fitted by the Lifshitz-Kosevich formula. 
				(c) Magnetoresistance $(\rho_{\mathrm{xx}}(B)-\rho_{\mathrm{xx}}(0$~$\mathrm{T}))/\rho_{\mathrm{xx}}(0$~$\mathrm{T})$ taken with rotating the magnetic field direction at $4.2$~K.
				The field angle $\theta$ is measured from the normal direction.
				(d) Field angle dependence of the oscillation frequencies obtained from the FFT spectra at $4.2$~K.
				All the peaks follow a ${\cos\theta}^{-1}$ dependence (dotted curves) rather than being constant (dashed curves).
			}
			\label{fig4}
		\end{center}
		\newpage
	\end{figure}
	The SdH oscillations appear at high fields below 40~K, where the Eu$^{2+}$ spin moments are oriented along the out-of-plane direction and band structures for the FM phase are expected to be realized (see Supplementary Materials \cite{Supplement}).
	Figure~4(a) plots the oscillatory component of the resistivity $\Delta \rho(B)$ as a function of $1/B$ involved at various temperatures.
	Apparently, there is more than one type of frequency.
	As shown in Fig.~4(b), the fast Fourier transform (FFT) of $\Delta \rho$ yields three major frequencies at $B_\mathrm{F}^\alpha=247$~T, $B_\mathrm{F}^\beta=372$~T, and $B_\mathrm{F}^\gamma=583$~T and their harmonics.
	The obtained frequencies suggest the presence of three Fermi pockets ($\alpha$, $\beta$, $\gamma$) perpendicular to the $c^\ast$ axis.
	As summarized in Table~1, the Fermi surface cross-section areas $A_\mathrm{F}$ perpendicular to the applied magnetic field are calculated to be $2.3$, $3.6$, and $5.6\times 10^{-2}$~\AA$^{-2}$, using the Lifshitz-Onsager relation, $B_\mathrm{F}=\frac{\hbar}{2\pi e}A_\mathrm{F}$ with the electron charge $e$ and the reduced Planck constant $\hbar$.
	These cross-sectional areas are typically small, taking only $\sim 2$~\% of the entire BZ on the $k_x$-$k_y$ plane.
	The FFT amplitudes of these three peaks gradually decrease with elevating temperature.
	As confirmed in the inset, their temperature dependence can be suitably fitted using the standard Lifshitz-Kosevich formula with thermal damping factors
	\begin{align}
		\Delta R(T)\propto  \frac{2 \pi^{2} k_{B} T / \hbar \omega_{c}}{\sinh \left(2 \pi^{2} k_{B} T / \hbar \omega_{c}\right)},
	\end{align}
	where $k_B$ is Boltzman constant and $\omega_{c}=\frac{eB}{m^\ast}$ the cyclotron frequency.
	The effective masses $m^\ast$ are estimated as small as about 0.11$m_0$ for all three peaks with the free electron mass $m_0$, as listed in Table~1.
	Importantly, this value is comparable to or even smaller than one reported in many other topological semimetals, e.g., 0.049$m_0$ (bulk) \cite{Uchida2017} and 0.27$m_0$ (surface) \cite{Nishihaya2019} in Cd$_3$As$_2$, 0.07$m_0$ (bulk) \cite{Zhang2017} and 0.50$m_0$ (surface) \cite{Nair2020} in TaAs, and 0.04-0.18$m_0$ (bulk) \cite{Matusiak2017} in ZrSiS.
	The Fermi velocity $v_\mathrm{F}=\frac{\hbar}{m^\ast}\sqrt{\frac{A_\mathrm{F}}{m^\ast}}$ is also estimated as large as $\sim4\times10^5$~ms$^{-1}$, similar to other topological semimetals \cite{Uchida2017,Nishihaya2019,Zhang2017,Nair2020,Matusiak2017}.
	More detailed analysis of the SdH oscillations indicates that the mobility is estimated as high as about $200$~cm$^2$V$^{-1}$s$^{-1}$ (for details see Supplementary Materials \cite{Supplement}).
	
	\begin{table}[htbp]
		\caption{Parameters for the three Fermi pockets extracted from the SdH oscillations.}
		\begin{ruledtabular}
			\begin{tabular}{ccccc}
				carrier & $B_\mathrm{F}$(T)& $A_\mathrm{F}$ (\AA$^{-2}$)& $m^\ast$ ($m_0$)& $v_\mathrm{F}$ ($10^5$~ms$^{-1}$) \\ \hline 
				$\alpha$ & 247 & 0.023 & 0.109&3.66 \\
				$\beta$& 372 & 0.036 & 0.113& 4.39 \\
				$\gamma$& 566 & 0.056 & 0.117& 5.24\\
			\end{tabular}
		\end{ruledtabular}
	\end{table}
	
	Figure~4(c) shows field-angle dependence of the SdH oscillations taken at $4.2$~K.
	The magnetoresistance is significantly suppressed with rotation of the field direction from the out-of-plane ($\theta=0^\circ$, $B \parallel c^\ast$) to the in-plane ($\theta=90^\circ$, $B \parallel a$-$b$) direction.
	In addition, peaks and valleys of the oscillations shift to higher fields upon increasing $\theta$.
	As confirmed in Fig.~4(d), although the oscillation amplitude becomes small and it is difficult to follow the frequency at higher angles ($\theta>40^\circ$), all frequencies extracted from the FFT spectra can be fitted to $1/\cos\theta$, indicating that the quantum transport observed in these EuSb$_2$ films originates from the 2D electronic structure.
	
	One possible origin of the 2D field-angle dependence is that EuSb$_2$ hosts a 2D bulk Fermi surface structure.
	However, most of the calculated Fermi surfaces are three-dimensional (3D), except only one 2D-like Fermi surface with low Fermi velocity (for details see Supplementary Materials \cite{Supplement}).
	Another possible origin is the quantum confinement effect.
	When the film thickness is comparable to the de Broglie wavelength of the carriers, the 3D bulk state is confined to form a 2D quantum-well state.
	However, the film thickness of 250~nm is considered large enough to maintain the 3D state, because even in Dirac semimetal Cd$_3$As$_2$ with extremely low carrier density and large Fermi velocity exhibits a 3D state above 100~nm film thickness \cite{Uchida2017}.
	A third possibility is surface states of EuSb$_2$.
	By fitting $\Delta \rho_\mathrm{xx}$ using the LK formula, we obtain nontrivial Berry Phases for the $\beta$ and $\gamma$ pockets (for details see Supplementary Materials \cite{Supplement}).
	These bands may be non-trivial surface states, which are likely to be protected by the combination of two-fold screw and time reversal symmetry.
	Another possibility is that they are trivial surface states formed on the $(001)$ polar surface, which can also obtain non-trivial phases with the Rashba splitting.
	In any case, it is unobvious why the 2D conduction state is much more dominant than the 3D state and further studies are needed to determine its origin.
	
	\section{CONCLUSION}
	In summary, we have fabricated single-crystalline EuSb$_2$ films by molecular beam epitaxy and have studied quantum transport at high fields.
	First-principles calculations have demonstrated that EuSb$_2$ hosts topological nodal lines protected by nonsymmorphic symmetry, which remains preserved even under AFM ordering.
	Effective masses extracted from multiple SdH oscillations are fairly small, which is characteristic of topological semimetals.
	The 2D field-angle dependence of the SdH oscillations suggests the possibility of surface quantum transport in EuSb$_2$.
	Our finding of the new magnetic TNLSM and observation of quantum transport will stimulate further investigations of exotic quantum transport as represented by Weyl orbit and topological phase transitions in magnetic topological semimetals.
	
	\section{ACKNOWLEDGMENTS}
	We thank M.-T. Huebsch and T. Yu for fruitful discussions.
	This work was supported by JST PRESTO No. JPMJPR18L2 and CREST Grant No. JPMJCR16F1, Japan and by Grant-in-Aids for Scientific Research on Innovative Areas No. 19H05825, Scientific Research (S) No. 16H06345, and Scientific Research (B) No. JP18H01866 from MEXT, Japan.

\end{document}


\title{Supplementary Materials for
	Quantum transport observed in films of magnetic topological semimetal EuSb$_2$
	}
	
	\author{Mizuki Ohno}
	\affiliation{Department of Applied Physics, University of Tokyo, Tokyo 113-8656, Japan}
	\affiliation{Quantum-Phase Electronics Center (QPEC), University of Tokyo, Tokyo 113-8656, Japan}
	\author{Masaki Uchida}
	\email[Author to whom correspondence should be addressed: ]{m.uchida@phys.titech.ac.jp}
	\affiliation{Department of Applied Physics, University of Tokyo, Tokyo 113-8656, Japan}
	\affiliation{Quantum-Phase Electronics Center (QPEC), University of Tokyo, Tokyo 113-8656, Japan}
	\affiliation{PRESTO, Japan Science and Technology Agency (JST), Tokyo 102-0076, Japan}
	\affiliation{Department of Physics, Tokyo Institute of Technology, Tokyo 152-8550, Japan}
	\author{Ryosuke Kurihara}
	\affiliation{The Institute for Solid State Physics (ISSP), The University of Tokyo, Kashiwa, Chiba 277-8581, Japan}
	\affiliation{RIKEN Center for Emergent Matter Science (CEMS), Wako 351-0198, Japan}
	\author{Susumu Minami}
	\affiliation{Department of Physics, University of Tokyo, Tokyo 113-8656, Japan}
	\author{\\Yusuke Nakazawa}
	\affiliation{Department of Applied Physics, University of Tokyo, Tokyo 113-8656, Japan}
	\affiliation{Quantum-Phase Electronics Center (QPEC), University of Tokyo, Tokyo 113-8656, Japan}
	\author{Shin Sato}
	\affiliation{Department of Applied Physics, University of Tokyo, Tokyo 113-8656, Japan}
	\affiliation{Quantum-Phase Electronics Center (QPEC), University of Tokyo, Tokyo 113-8656, Japan}
	\author{Markus Kriener}
	\affiliation{RIKEN Center for Emergent Matter Science (CEMS), Wako 351-0198, Japan}
	\author{\\Motoaki Hirayama}
	\affiliation{Department of Applied Physics, University of Tokyo, Tokyo 113-8656, Japan}
	\affiliation{RIKEN Center for Emergent Matter Science (CEMS), Wako 351-0198, Japan}
	\author{Atsushi Miyake}
	\affiliation{The Institute for Solid State Physics (ISSP), The University of Tokyo, Kashiwa, Chiba 277-8581, Japan}
	\author{Yasujiro Taguchi}
	\affiliation{RIKEN Center for Emergent Matter Science (CEMS), Wako 351-0198, Japan}
	\author{\\Ryotaro Arita}
	\affiliation{Department of Applied Physics, University of Tokyo, Tokyo 113-8656, Japan}
	\affiliation{RIKEN Center for Emergent Matter Science (CEMS), Wako 351-0198, Japan}
	\author{Masashi Tokunaga}
	\affiliation{The Institute for Solid State Physics (ISSP), The University of Tokyo, Kashiwa, Chiba 277-8581, Japan}
	\affiliation{RIKEN Center for Emergent Matter Science (CEMS), Wako 351-0198, Japan}
	\author{Masashi Kawasaki}
	\affiliation{Department of Applied Physics, University of Tokyo, Tokyo 113-8656, Japan}
	\affiliation{Quantum-Phase Electronics Center (QPEC), University of Tokyo, Tokyo 113-8656, Japan}
	\affiliation{RIKEN Center for Emergent Matter Science (CEMS), Wako 351-0198, Japan}
	
	\maketitle
	\section{Structural characterization}
	Further x-ray diffraction (XRD) characterization was performed for elucidating the crystallinity and in-plane orientation of the EuSb$_2$ films.
	As shown in the out-of-plane XRD ${\theta}$-${2\theta}$ scan in Fig.~S1(a), only sharp reflections from the (00$l$) lattice plane of EuSb$_2$ are observed without any impurity phases.
	A rocking curve in Fig.~S1(b) for the (003) EuSb$_2$ film peak is very sharp with a full width at half maximum of $0.03$~degrees, ensuring high crystallinity of the obtained film.
	The in-plane XRD reciprocal space map in Fig.~S1(c) indicates that the $(020)$  EuSb$_2$ film plane and the $(\bar{3}30)$  Al${_{2}}$O${_{3}}$ substrate plane are orthogonal. Namely, the EuSb$_2$ $b$ axis is parallel to the Al$_2$O$_3$ $c$ axis, as illustrated in Figs.~S1(b) and S1(c).
	\begin{figure}
		\begin{center}
			\includegraphics*[bb=0 0 364 448,width=8.6cm]{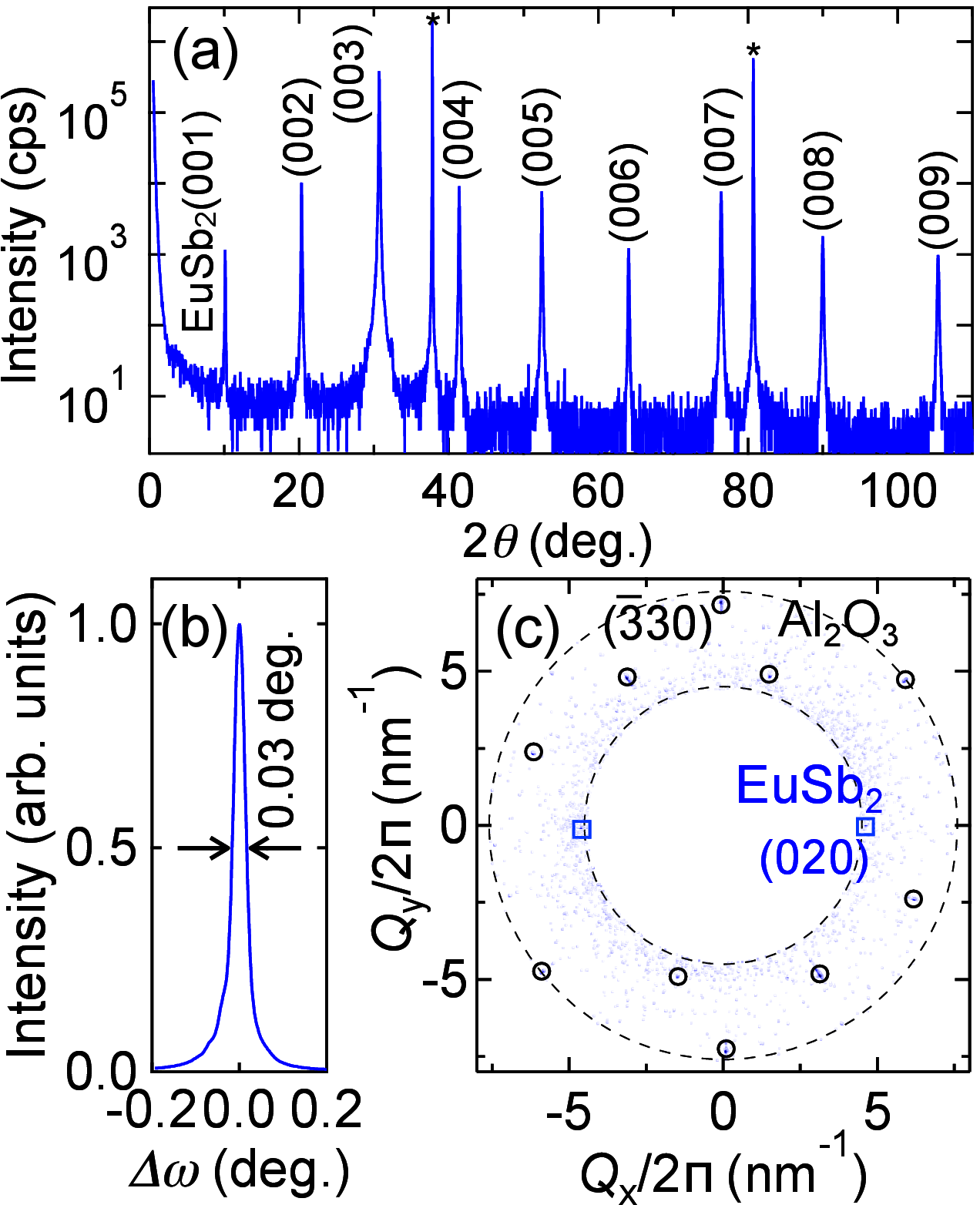}
			\caption{
				(a) XRD ${\theta}$-${2\theta}$ scan of a EuSb$_2$ film grown on Al${_{2}}$O${_{3}}$ (11$\bar{2}$0) substrate. Al$_2$O$_3$ substrate peaks are marked with an asterisk. (b) Rocking curve of the (003)  EuSb$_2$ film peak. (c) In-plane XRD reciprocal space mapping.
			}
			\label{figS1}
		\end{center}
	\end{figure}
	\clearpage
	\section{Band structure calculation}
	The antiferromagnetic (AFM) ordering, oriented along the [010] direction, belongs to the magnetic space group $P_a2_1/m$ and does not break the two-fold screw symmetry ($2_1/m$) along the $b$ axis.
	It is thus expected that the nodal lines on the $k_y=\pi/b$ plane stably exist in the AFM phase.
	In the band structure calculated for the AFM phase in Fig.~S2(a), band crossing points are confirmed on the high-symmetry points Z, A, E, and D, which are connected by the nodal lines on the $k_y=\pi /b$ plane as also presented in the main text.
	
	Ferromagnetic (FM) ordering aligned along the [001] direction breaks the two-fold screw symmetry.
	Hence, it is expected that the nodal lines on the $k_y=\pi/b$ plane is unstable in the FM phase. 
	The band structure of the FM phase with $B\parallel c^*$ is shown in Fig.~S2(b).
	The band crossing points on the high-symmetry points of Z, A, E, and D are gapped and the nodal lines on the $k_y=\pi/b$ plane have disappeared.
	Four bands, marked in purple, blue, green, and pink cross the Fermi level in the FM phase.
	To further examine these bands crossing of the Fermi level, we plot their Fermi surface structures with Fermi velocity in Fig.~S2(c).
	Band~1 has a cylinder-like Fermi surface, where two-dimensional (2D) field-angle dependence is expected in the SdH oscillations.
	However, there are other Bands~2-4 with comparable or even lower Fermi velocity, where complicated three-dimensional (3D) Fermi surfaces are realized.
	Therefore, if the observed SdH oscillation is ascribed to Band~1, oscillations from the other bands with complex 3D field-angle dependence should be also observed.
	In addition,the Fermi velocity estimated from the observed SdH oscillation is one order of magnitude lower than those calculated for Band~1.
	Even considering correlation effects, reduction of the calculated Fermi velocity is limited since the dispersive bands crossing $E_\mathrm{F}$ mainly derive from the Sb 5$p$ orbitals. 
	\begin{figure}[h]
		\begin{center}
			\includegraphics*[bb=0 1 720 420,width=15cm]{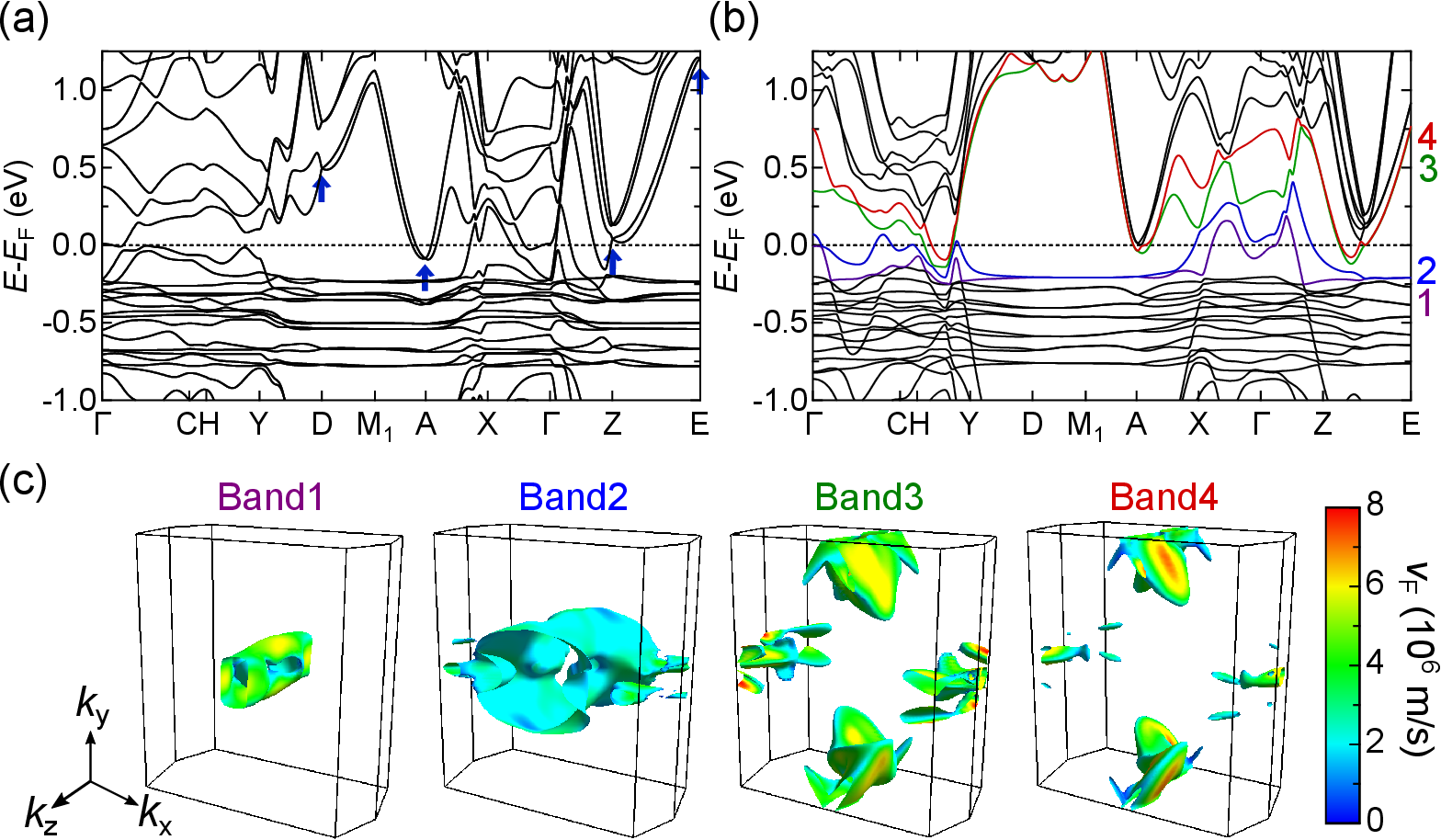}
			\caption{
				(a) First-principles band structure of EuSb$_2$ calculated for AFM ordering. Blue arrows represent the band crossing points protected by the two-fold screw symmetry. (b) Band structure calculated for FM ordering along the $c^*$ axis. (c) Fermi surfaces in the FM phase, colored by the Fermi velocity. Four bands in (b), plotted by purple, blue, green, and pink, correspond to bands 1-4.
			}
			\label{figS4}
		\end{center}
	\end{figure}
	\clearpage
	\section{Transport and magnetic properties for the in-plane field}
	Figure~S3(a) shows magnetization curves measured at various temperatures. The magnetization at $1.4$~K exhibits a clear spin-flop transition at $2.4$~T, corresponding to reorientation of the spin moments perpendicular to the applied field.
	After the spin-flop transition, the magnetization at $1.4$~K increases approximately linearly with fields and then saturates at $13.7$~T, indicating transition from the spin-flop AFM phase to the FM one.
	In addition, this field is almost same as in the out-of-plane magnetic field, reflecting the isotropic feature of the Eu$^{2+}$ spin moments. Temperature dependence of magnetization also shows a clear kink at $T_\mathrm{N}$.
	Figure~S3(b) presents in-plane magnetic field sweeps of the resistivity at various temperatures.
	As more clearly seen in its first derivative in Fig.~S3(c), the minimum shifts to lower fields with increasing temperature and then disappears below $T_\mathrm{N}$, also corresponding to the transition from the spin-flop AFM phase to the FM one.
	The maximum around $2.4$~T observed below $T_\mathrm{N}$ is interpreted as the spin-flop transition. 
	Figure S3(d)~summarizes the in-plane $B$-$T$ phase diagram as determined by the above transport and magnetization measurements, which is typical of in-plane antiferromagnets.
	\begin{figure}[h]
		\begin{center}
			\includegraphics*[bb=0 0 630 435,width=15cm]{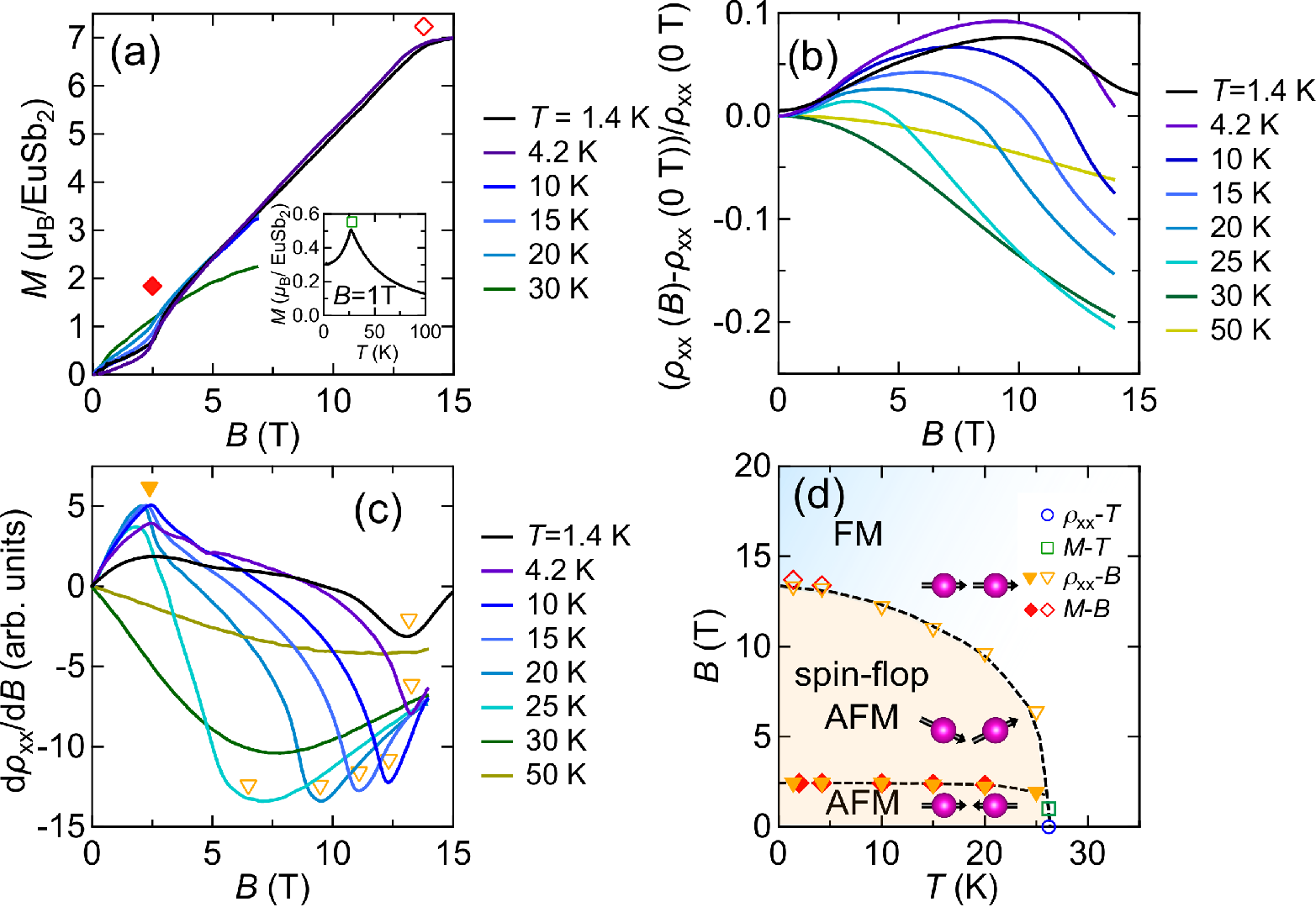}
			\caption{
				(a) Magnetization curve taken at various temperatures, where the spin-flop transition and saturation fields are marked by filled and open diamonds.
				The inset shows temperature dependence of the magnetization $M$, taken with applying an in-plane field of $1.0$~T.
				It exhibits a clear kink at $26.2$~K, as marked by a square.
				(b) Magnetoresistance $(\rho_{\mathrm{xx}}(B)-\rho_{\mathrm{xx}}(0$~$\mathrm{T}))/\rho_{\mathrm{xx}}(0$~$\mathrm{T})$ taken with sweeping the in-plane magnetic field $(B\parallel I)$ at various temperatures.
				(c) First derivative of $\rho_{\mathrm{xx}}$, where the transition fields to the spin-flop AFM and forced FM phases are confirmed as marked by filled and open triangles.
				(d) Magnetic phase diagram obtained for the in-plane field. The dashed curves represent the phase boundaries. Black arrows are schematic illustration of the Eu spin moments.
			}
			\label{figS3_figR3}
		\end{center}
	\end{figure}

	\clearpage
	\section{Anomalous Hall effect}
	Figure~S4(a) shows the Hall resistivity $\rho_{\mathrm{yx}}$ taken at different temperatures.
	At $300$~K, $\rho_{\mathrm{yx}}$ is almost linear with a positive slope, which indicates holes are majority carriers, on the other hand, $\rho_{\mathrm{yx}}$ starts to deviate from the linear trend with decreasing temperature.
	In addition, $\rho_{\mathrm{yx}}$ below $T_\mathrm{N}$ exhibits a cusp-like behavior around $14$~T, which cannot be fitted by the simple two-carrier model.
	In order to confirm the presence of anomalous Hal effect, we convert it to the Hall conductivity $\sigma_{\mathrm{yx}}$.
	As shown in Fig.~S4(b), $\sigma_{\mathrm{yx}}$ also starts to deviate from the linear trend with decreasing temperature. Below $T_\mathrm{N}$, this nonmonotonic behavior becomes linear (with fine intercept at the zero field) above the saturation field and also shows a bump and dip structure below the saturation field.
	It can be thus concluded that there are clear anomalous Hall signals above and below the saturation field.
	This bump and dip structure is similar to one observed in other Eu-based magnetic compound, which is ascribed to the change of electronic bands or Weyl nodes induced by the Zeeman splitting during the spin canting process of Eu$^{2+}$ spin \cite{Takahashi2018}.
	Although contribution of possible multi-carriers makes it difficult to analyze these complicated signals more quantitatively, the clear anomalous Hall effect indicates the strong coupling between the Eu localized spins and the conduction carriers.
	\begin{figure}[h]
		\begin{center}
			\includegraphics*[bb=0 0 638 214,width=15cm]{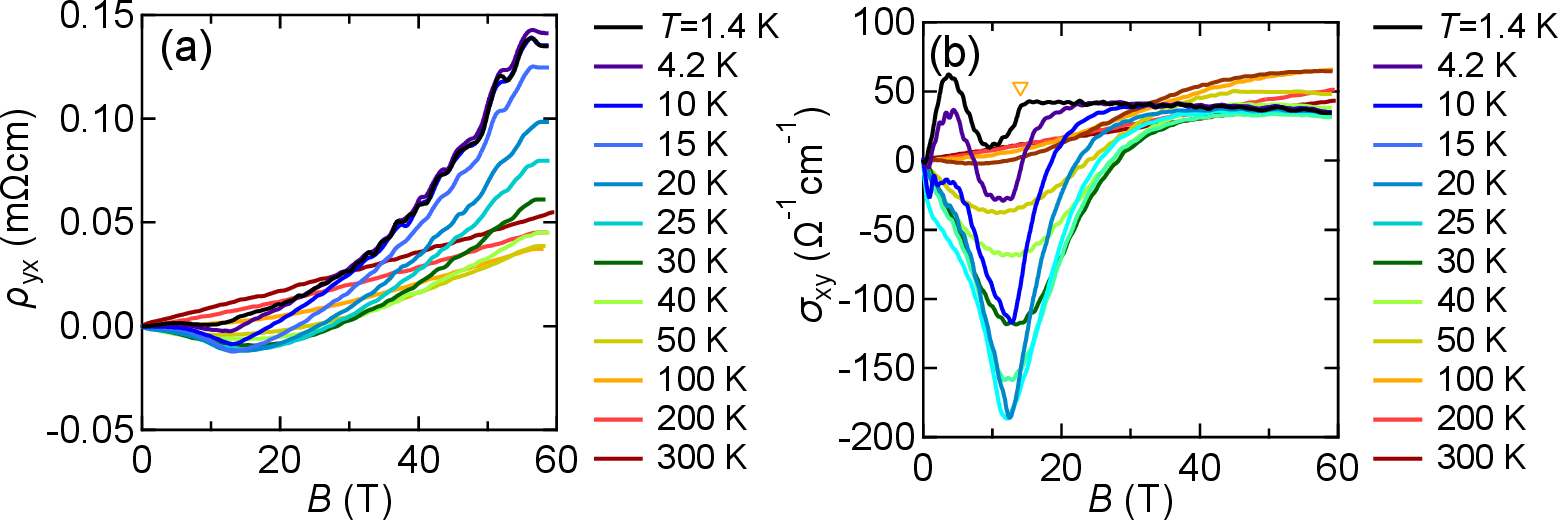}
			\caption{
				(a) Hall resistivity $\rho_{\mathrm{yx}}$ and (b) Hall conductivity $\sigma_{\mathrm{yx}}$ taken with sweeping the out-of-plane magnetic field at various temperatures.
				The saturation field is clearly confirmed as marked by a triangle.
			}
			\label{figS2_figR5}
		\end{center}
	\end{figure}
	
	\clearpage
	\section{Berry Phase estimation}
	The observed oscillatory component of the resistivity $\Delta \rho_\mathrm{xx}$ is fitted using the LK formula
	\begin{align}
		\Delta \rho_\mathrm{xx} \propto  R_T R_D R_S \cos\left[ 2 \pi\left(\frac{B_\mathrm{F}}{B}+\gamma-\delta \right)\right],
	\end{align}
	where the thermal damping factor $R_T=\alpha/\sinh{(\alpha)}$ with $\alpha=2\pi^2k_BT/\left(\hbar \omega_c\right)$ and the Dingle damping factor $R_D=\exp\left[-2\pi^2k_BT_D/\left(\hbar\omega_c\right)\right]$ with the Dingle temperature $T_D$.
	$R_S=\cos(\pi gm^\ast/2m_0)$ is the spin damping factor with the Lande factor $g$.
	In the case of 2D oscillations, $\delta$ is zero and the phase factor $\gamma$ is related with the Berry phase $\mathit{\Phi}_\mathrm{B}$ by $\gamma=1/2-\mathit{\Phi}_\mathrm{B}/2\pi$ \cite{Mikitik1999}.

	To examine the Berry phase effects, we first fit $\Delta \rho_\mathrm{xx}$ measured at $T=1.4$~K by the generalized LK formula with three oscillation frequencies, as shown in Fig.~S5(a).
	The obtained Berry phases are $\mathit{\Phi}_\mathrm{B}^\alpha=0.01\pi$, $\mathit{\Phi}_\mathrm{B}^\beta=1.27\pi$, and $\mathit{\Phi}_\mathrm{B}^\gamma=1.52\pi$, indicating that the Fermi pocket $\alpha$ has a trivial phase and the other two pockets $\beta$ and $\gamma$ have nontrivial Berry phases.
	Deviation from the ideal nontrivial phase of $\pi$ may be ascribed to non ideal linear dispersion as reported in other topological semimetals such as NbP \cite{Wang2016}.
	As shown in Fig.~S5(b), similar fitting results are obtained also for the data at $T=10$~K.
	Extracted scattering times $\tau_q$ are about 10$^{-14}$~s and mobilities estimated by $\mu_q=e\tau_q/m^\ast$ are about $200$~cm$^2$V$^{-1}$s$^{-1}$ below 10~K, also reflecting high quality of the film.
	\begin{figure}[b]
		\begin{center}
			\includegraphics*[bb=0 0 481 440,width=13cm]{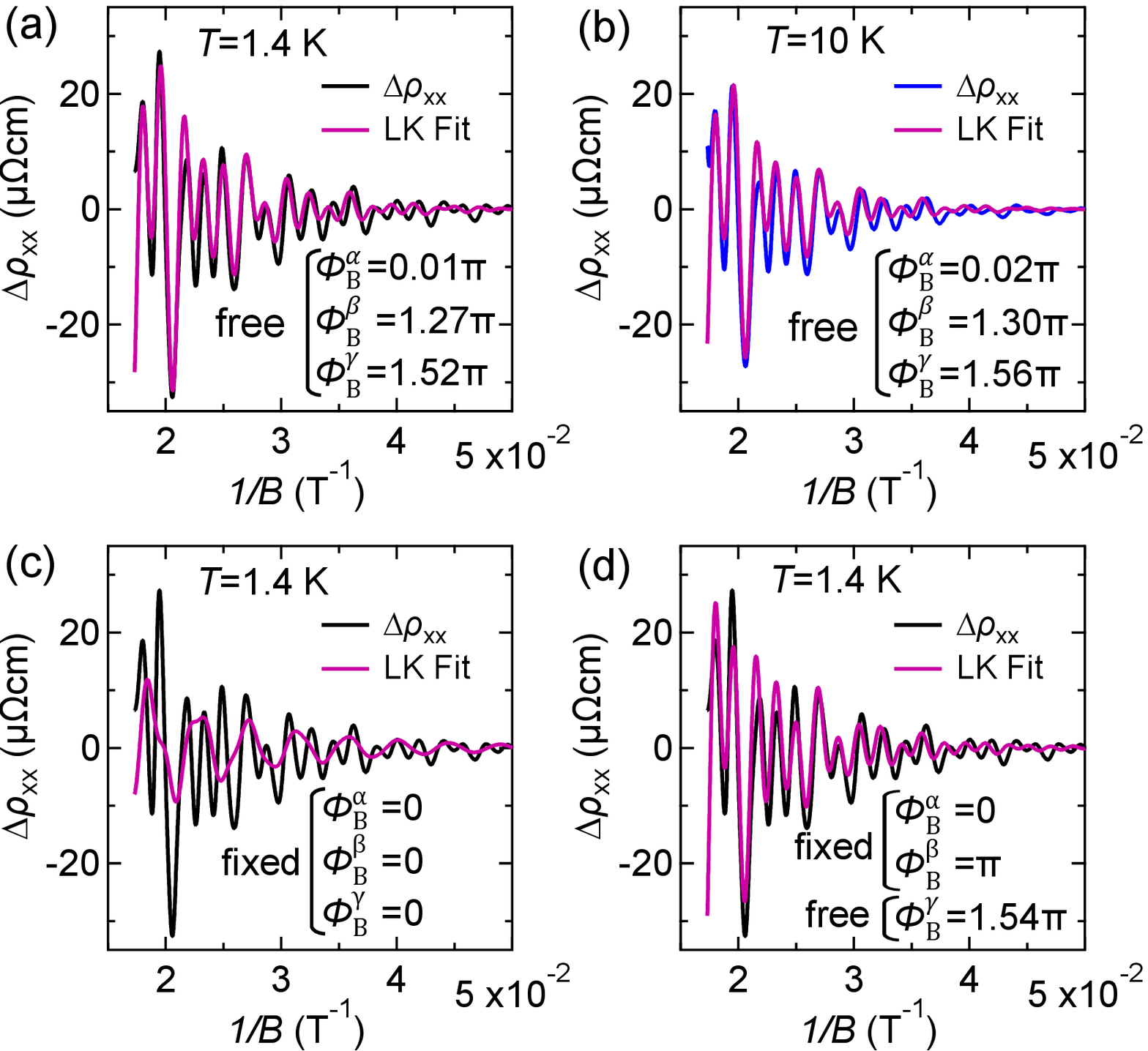}
			\caption{
				(a) SdH oscillation part $\Delta \rho_\mathrm{xx}$ and the fit to LK formula with three frequencies ($B_\mathrm{F}^\alpha$, $B_\mathrm{F}^\beta$, and $B_\mathrm{F}^\gamma$), plotted against $1/B$ at $T=1.4$~K.
				(b) $\Delta \rho_\mathrm{xx}$ and the fit to LK formula at $T=10$~K.
				Other trial fittings (c) with fixed parameters of $\mathit{\Phi}_\mathrm{B}^\alpha=0$, $\mathit{\Phi}_\mathrm{B}^\beta=0$, and $\mathit{\Phi}_\mathrm{B}=0$, and (d) with fixed parameters of $\mathit{\Phi}_\mathrm{B}^\alpha=0$ and $\mathit{\Phi}_\mathrm{B}^\beta=\pi$ and a free parameter of $\mathit{\Phi}_\mathrm{B}^\gamma$.
			}
			\label{figS5_figR1}
		\end{center}
	\end{figure}

	To check robustness of these fitting results, we attempted to fit the oscillations using different phase factors or Berry phases.
	When all the parameters of the Berry phases $\mathit{\Phi}_\mathrm{B}^\alpha$, $\mathit{\Phi}_\mathrm{B}^\beta$, and $\mathit{\Phi}_\mathrm{B}^\gamma$ are fixed to zero, the data cannot be fitted at all, as confirmed in Fig.~S5(c).
	On the other hand, for example, when $\mathit{\Phi}_\mathrm{B}^\alpha$ and $\mathit{\Phi}_\mathrm{B}^\beta$ are fixed to $0$ and $\pi$ and $\mathit{\mathit{\Phi}}_\mathrm{B}^\gamma$ is set as a free fitting parameter, the data can be well fitted with a nontrivial $\mathit{\Phi}_\mathrm{B}^\gamma$ value, as shown in Fig.~S5(d).
	From these fitting results, we suggest that the two Fermi pockets $\beta$ and $\gamma$ have nontrivial Berry phases.
	These $\beta$ and $\gamma$ bands may be non-trivial surface states, which is likely to be protected by the combination of two-fold screw and time reversal symmetry.
	Another possibility is that they are trivial surface states formed on the $(001)$ polar surface, which can also obtain non-trivial phases with the Rashba splitting.